# Features of the Electronic and Magnetic Properties of Heusler Alloys in the States of a Half-Metallic Ferromagnet and a Spin-Gapless Semiconductor

**V. V. Marchenkov, V. Yu. Irkhin, and Yu. A. Perevozchikova**
*M.N. Mikheev Institute of Metal Physics, Ural Branch, Russian Academy of Sciences, Ekaterinburg, 620108 Russia*
*\*e-mail: march@imp.uran.ru*

The review treats Heusler alloys that display distinctive functional properties, including shape-memory behavior and magnetocaloric effects. Particular emphasis is placed on Heusler systems in which half-metallic ferromagnetism and spin-gapless semiconductor state are realized. Although these compounds are crystallographically rather "ordinary," peculiarities of their electronic structure and magnetic state lead to unconventional kinetic and magnetic properties. Their magnetic and transport characteristics are highly sensitive to external stimuli, and changes in alloy composition or external parameters can induce transitions between the states considered. This tunability provides further opportunities for controlling the electronic and magnetic properties of Heusler alloys and for exploiting them in applications such as spintronics and micro- and nanoelectronics.

Keywords: Heusler alloys, half-metallic ferromagnets, spin-gapless semiconductors

## INTRODUCTION

In 1903, the German chemist F. Heusler discovered a new type of compound, which was later named after him – Heusler alloys [1]. To date, more than one and a half thousand different Heusler compounds are known to exhibit unique physical properties (e.g., reviews [2-5] and references therein). In our opinion, the most interesting of them are Heusler alloys with the shape memory effect (SME) and the magnetocaloric effect (MCE), half-metallic ferromagnets (HMF), and spin-gapless semiconductors (SGS) [6, 7].

Intensive studies of MCE began in the late 90s and early 2000s, when it was shown [8] that a giant MCE occurs in gadolinium and gadolinium-based compounds $Gd_5(Si_2Ge_2)$. High values of magnetic anisotropy near room temperature were also observed in $La(FeSi)_{13}$ and MnAs-based compounds [9]. MCE was observed in Heusler alloys and $Ni_2MnGa$ as one of its classical representatives of compounds, which exhibit both SME, which is controlled by magnetic field and temperature, and MCE (see, e.g., [10]). Currently, a large number of researchers worldwide, including Russia, are studying shape memory and caloric effects. Russian groups perform this research in Chelyabinsk (CSU) [11-16], Moscow (IRE RAS, MISIS, MSU, etc.) [11-21], Makhachkala (DFRC RAS) [15, 20-23], Donetsk (DonIPE, DonNU) [20, 21],

Ufa (Ufa University of Science and Technology, IMSP RAS) [12, 16, 17], Yekaterinburg (IMP UB RAS, UrFU, IEP UB RAS) [17, 24-28], Tomsk (TSU, etc.) [29-31], and other research centers.

An equally interesting area of research is the study of the states of HMF and SGS, including in Heusler alloys [2, 3, 5]. The HMF state occurs when an energy gap is observed in the electronic structure near the Fermi level for one spin direction, and a high density of states is observed for the other spin. This state was predicted for NiMnSb [32] and PtMnSb [33, 34] by de Groot et al. and for $Co_2MnSn$ by Kubler et al. [35]. The HMF state has been the subject of intensive theoretical [36] and experimental [5, 6] studies.

The SGS state was predicted in the theoretical work of Wang [37]. This state resembles the HMF state in many ways; the only difference is that instead of a high density of states for one spin, a zero gap is observed. This zero gap phenomenon occurs in "classical" gapless semiconductors intensively studied by Tsidilkovski and colleagues [38, 39].

In addition to the significant fundamental interest, the comprehensive study and implementation of HMF and SGS states is promising for their potential application in spintronics due to the high degree of spin polarization in these materials, even at room temperature [5]. In SGS materials, almost no energy is required to excite an electron from the valence band to the conduction band. In this case, the excited electrons are 100% spin-polarized for excitation energies up to the gap band energy of another spin channel. Even more interestingly, holes can also be 100% spin-polarized. It is important to note that all related properties can be adjusted by external influences such as pressure, electric and magnetic fields, electromagnetic radiation, impurities, etc. [37].

A large number of publications have been devoted to the study of HMF and SGS Heusler alloys, which mainly consist of "classical" full $X_2YZ$ and half-Heusler XYZ compounds. Here, X and Y are typically transition metals and Z are $s$ and $p$ elements from the periodic table. Nevertheless, an increasing number of new works have appeared in this field, particularly concerning quaternary Heusler compounds [40-42], all-$d$ systems [43-45], and even exotic $Z_2XY$ Heusler alloys [46-48]. Therefore, it is necessary to provide an overview of the existing concepts of HMF and SGS states in Heusler alloys, including recent publications on this topic. Another objective of this review is to discuss the features of electronic transport and magnetic properties in HMF and SGS states, new "exotic" Heusler compounds, and potential avenues for future research in this area.

## HALF-METALLIC FERROMAGNETS

Magnetic materials exhibit two types of electronic state: spin "up" and spin "down." The electronic kinetic properties of such materials can be modelled as having two conduction channels, corresponding to current carriers with "up" and "down" spins. In HMF materials, over a wide temperature range (much lower than the "gap temperature"), charge carriers with "down" spins freeze out and electronic transport is mainly determined by the "up" spin conduction channel. This can lead to a spin polarization of current carriers close to 100%.

In this case, scattering of conduction electrons by magnons is only possible through two-magnon scattering processes [49]. This scattering mechanism results in a power-law contribution to the temperature dependence of the electrical resistivity, $\rho \sim T^n$, where $7/2 < n < 9/2$. The authors of the experimental work [50] on the electrical resistivity $\rho(T)$ of polycrystalline Heusler alloys $Co_2FeSi_{1-x}Z_x$ (where Z = Al, Ge, Ga, Sn, or In; $x = 0$ or 0.5) observed a contribution to resistivity proportional to $T^{9/2}$. A dependence of $\rho \sim T^4$ was observed in monocrystalline $Co_2FeSi$ [51]. More recently, in the $Co_2FeAl$, $Co_2FeSi$, and $Co_2FeGe$ Heusler alloys, a power-law dependence of electrical resistivity $\sim T^n$ with $n$ in the range $3.5 \leq n \leq 4$ was observed in the temperature range of 40 to 65 K. According to the authors [50-52], these observed dependencies are manifestations of two-magnon scattering processes.

The formation of a half-metallic state in $X_2MnZ$ and/or $XMnZ$-type Heusler alloys can be explained as follows [32, 34, 35, 53]. Neglecting the hybridization of the atomic states of the X and Z elements reveals that the manganese $d$-states exhibit a substantial energy gap between the bonding and anti-bonding states. Due to strong intra-atomic (Hund's rule) exchange in the ferromagnetic state, the subbands with opposite spins of the manganese ions are significantly separated in energy. One of these spin subbands approaches the $p$-states of the Z ligand, thus leading to the partial or complete disappearance of the corresponding gap due to $p$–$d$ hybridization. Meanwhile, the second subband retains its energy gap, into which the Fermi level can enter under certain conditions, ensuring a half-metallic ferromagnetic (HMF) state.

In Heusler HMF alloys, "strong" and/or "weak" magnetism can occur, with different magnetic moment values depending on whether a localized or collectivized magnetic state is realized, or a combination of the two. The authors of [54] calculated the electronic structure and demonstrated that Heusler XYZ compounds (X = Fe, Co, Ni, Rh, Ir, Pd, or Pt; Y = Ti, V, Zr, Hf, Nb, or Ta; Z = In, Sn, or Sb) with a gap can exist in specific magnetic states. Co- and Ni-based alloys predominantly demonstrate collectivized magnetism, whereas Ti-, V-, and Fe-based compounds can be in either a localized or collectivized magnetic state, or in a state of coexistence.

It is known that the Slater-Pauling rule is fulfilled in Heusler HMF alloys due to the presence of a gap, establishing a relationship between the number of valence electrons $z$ and the magnetic moment $M$ [55]. This allows one to predict the total magnetic moment, $M = z - 24$. According to this rule, the total magnetic moment in Heusler alloys with $z = 24$ must be zero. A recent paper [46] reported the existence of a special class of Heusler compounds half consisting of $p$-elements. This class of compounds includes $Al_2MnCu$, which has 24 valence electrons. According to [46], this compound has a large magnetic moment of ~1.8 $\mu_B$/f.u. and a Curie temperature of $T_C \sim$ 315 K, and the number of valence electrons is $z$ = 24. To provide a preliminary explanation for the deviation from the Slater-Pauling rule, the authors of [46] propose a phenomenological model within the molecular orbital approach.

Another "unusual" class of Heusler compounds is all-$d$ alloys consisting entirely of $d$ elements [44]. The electronic structure and magnetic properties of equiatomic quaternary Heusler alloys (EQHA) based on Zn XX'YZn (where X = Fe, Ni, or Co; X' = V or Cr; and Y = Nb, Ti, Zr, or Hf) were studied using the first-principles density functional calculations [44]. It has been demonstrated that 100% spin polarization can occur in these compounds. According to [44], NiVTiZn and NiVZrZn alloys are HMFs, whereas FeVNbZn, CoCrTiZn, CoCrZrZn, and CoCrHfZn alloys are half-metallic ferrimagnets. The magnetic moments of these six compounds are primarily determined by the magnetic moments of the V and Cr atoms.

It is also worth noting that Heusler alloys can exhibit unique characteristics, as well as useful combinations of these. In particular, the authors of [56] present a comprehensive study of the structure and functional properties of $Co_2VGa_{1-x}Sbx$ ($x$ = 0, 0.25 or 0.5) Heusler alloys, in which gallium atoms are replaced with antimony. They demonstrate that compositions with $x$ = 0 and 0.25 retain half-metallic properties and 100% spin polarization. Therefore, $Co_2VGa_{0.75}Sb_{0.25}$ is a promising material for spintronics applications. At $x$ = 0.5, the alloy undergoes a martensitic transformation from the cubic $L2_1$ phase to the tetragonal $D0_{22}$ phase at a temperature of ~347 K. This transformation is important for potential applications as a shape memory material. $Co_2VGa_{0.5}Sb_{0.5}$ retains half-metallic properties despite an increase in the lattice parameter by 2%, which confirms the stability of its characteristics.

## SPIN-GAPLESS SEMICONDUCTORS

Spin-gapless semiconductors have the following features: (1) a small amount of energy is required to excite charge carriers; (2) fully spin-polarized charge carriers (electrons and/or holes) can be generated by light excitation or thermal activation; (3) charge carriers are fully spin-polarized; (4) under the influence of a magnetic field, polarized carriers can move to the edge of thin samples due to the Hall effect [57]. Not only electrons, but also holes can be 100% spin-polarized in spin-gapless semiconductors. These features lead to unique transport properties, i.e., the coexistence of high values of electrical resistivity, Curie temperature, magnetic moment, and a small anomalous Hall effect [58].

Full Heusler alloys can form various structures. Ordered structures can be found in the regular Heusler alloy $L2_1$ (Fm$\overline{3}$m) and the inverse Heusler alloy $XA$ (F$\overline{4}$3m), in which the X, Y, and Z atoms are located in distinct sublattices [3, 36]. Disordered structures can also form the Heusler alloys. Examples include B2 ($Pm\bar{3}m$), where the Y and Z atoms are evenly distributed, A2 ($Im\bar{3}m$), where all the X, Y, and Z atoms are evenly distributed, and other structures [3].

Among Heusler alloys, the $Mn_2CoAl$ compound, which is formed in the inverse Heusler alloy structure (*XA* structure), is of particular interest because the SGS state was experimentally and theoretically observed in it for the first time [58]. This compound exhibits high magnetic moments (2 $\mu_B$), a high Curie temperature (720 K), a high resistivity (residual resistivity ~440 $\mu\Omega$ cm), a low anomalous Hall coefficient (~22 S/cm at 2 K), concentrations of charge carriers (~$10^{20}$ $cm^{-3}$), and a Seebeck coefficient (~2 $\mu$V/K) [58]. If the $L2_1$ structure of a regular Heusler alloy forms in the $Mn_2CoAl$ and $Mn_2CoGa$ compounds, the magnetic moment remains at 2 $\mu_B$, and the spin-gapless semiconductor state transforms into a half-metallic ferromagnetic state [59]. However, if the B2 or β-Mn structure is observed, the SGS and HMF states do not form in $Mn_2CoAl$, $Mn_2FeAl$, and $Mn_2NiAl$, but rather a metallic density of electronic states is observed [60, 61]. In this case, the magnetic moment in the β-Mn structure of $Mn_2CoAl$ and $Mn_2NiAl$ decreases to 0.56 and 0.32 $\mu_B$/f.u., respectively, and an antiferromagnetic state with a zero magnetic moment is formed in the B2 structure. The residual electrical resistivity in $Mn_2CoAl$ remains high but is half that observed in *XA* (~242 $\mu\Omega$ cm) [58, 60]. Interestingly, the SGS state of the compound is preserved when deviating from the stoichiometry ($Mn_{1.8}Co_{1.2}Al$) [62].

Nowadays, Heusler quaternary alloys are studied intensively [63-65]. The electronic structure and properties of IrCoTiZ Heusler alloys (Z = Si, Sn, or Pb) have been investigated theoretically [65]. These compounds were found to be spin-gapless semiconductors with an energetically stable ferromagnetic phase, a total magnetic moment of 2 $\mu_B$, and a Curie temperature above 1000 K. The IrCoTiSi compound exhibits a strong optical response over a wide range of energies, making it an excellent candidate for optoelectronic applications. From the thermal electromotive force (EMF) data, the authors [65] determined that IrCoTiPb and IrCoTiSn exhibit *p*-type behaviour for both spin projections, whereas IrCoTiSi exhibits mixed behavior (i.e., *n*-type for spin down and *p*-type for spin up) [65].

The authors of [64] experimentally studied a quaternary CoFeMnSn alloy that was formed within the LiMgPdSn-type ($F\bar{4}3m$) Heusler structure. The magnetic moment value of 4.04 $\mu_B$/f.u. corresponds to the Slater-Pauling rule. The Curie temperature is 660 K, the residual resistivity is ~328 $\mu\Omega$ cm, the anomalous Hall coefficient at 2 K is 83.5 S/cm, and the Seebeck coefficient is 5.7 $\mu$V/K at room temperature. Analysis of the Hall effect and thermal EMF data revealed that holes are the dominant charge carriers.

Although the SGS state is most commonly observed in inverse Heusler alloys, the authors of [63] theoretically identified a nearly SGS state in FeVTaAl (type II SGS) and FeCrZrAl (type I SGS) alloys with a regular $L2_1$ Heusler structure and an integral magnetic moment at an equilibrium lattice constant [63]. Currently, various materials exhibiting spin-gapless semiconductor properties are being actively investigated, thus making it possible to select a compound with the necessary properties for practical use.

## CONCLUSIONS

Thus, we examined two important classes of Heusler alloys: half-metallic ferromagnets (HMFs) and spin-gapless semiconductors (SGSs). HMFs are characterized by the presence of a gap at the Fermi level for spin-down electronic states and the absence of a gap for spin-up charge carriers. The behavior of electronic transport in these systems can be modeled with two conduction channels. Spin-gapless semiconductors are a new class of quantum material with a unique spin-polarized band structure. They have a finite band gap only for carriers with one spin direction, and a zero or near-zero gap for carriers with the opposite direction, which makes them promising materials for spintronics devices.

Once again, we will outline the latest developments in Heusler alloy research. These trends include searching for and studying new quaternary and double half-Heusler alloys (such as XX'YZ and XX'YYZZ'), alloys consisting entirely of $d$-metals (known as all-$d$-metal Heusler compounds), new compounds (such as $Al_2MnCu$), synthesizing single crystals of Heusler alloys, and studying various effects such as rapid quenching from a melt, thermobaric high-pressure treatment, and high-energy particle irradiation and others. New electronic and magnetic effects include the occurrence of large thermal EMF, the observation of giant Nernst-Ettingshausen and anomalous Hall effects, metamagnetic transitions in extremely strong magnetic fields, etc.

Thus, new materials based on Heusler alloys and the novel arising phenomena can be used in spintronics and micro- and nanoelectronic devices.

## FUNDING

The work was supported by the state assignment of the Ministry of Science and Higher Education of the Russian Federation for the IMP of the Ural Branch of the Russian Academy of Sciences.